\documentclass[aps,prl,reprint]{revtex4-1}
\usepackage{graphicx}
\begin{document}
\title{Periodic Variation of Stress in Sputter \\ Deposited Si/WSi$_2$ Multilayers}
\author{Kimberley MacArthur$^{1,a}$}
\author{Bing Shi$^1$}
\author{Ray Conley$^{1,2,b}$}
\author{Albert T. Macrander$^{1,c}$}
\affiliation{$^1$Advanced Photon Source, Argonne National Laboratory, Argonne, Illinois, 60439, USA}
\affiliation{$^2$NSLS-II, Brookhaven National Laboratory, Upton, New York 11973, USA} 
\date{\today}
\begin{abstract}
   A tension increment after sputter  deposition of   1  nm of WSi$_2$ onto sputtered  Si  was observed at low Ar gas pressures. Wafer curvature data on multilayers were found to have a periodic variation corresponding to the multilayer period, and this permitted   statistical analyses to improve the sensitivity to small stresses. The observation of tension instead of compression in the initial stage of growth is new , and a model invoking surface rearrangement is invoked. The data also bear on an unusual surface  smoothing  phenomena for  sputtered Si surfaces caused by the  sputter  deposition of WSi$_2$ . We furthermore report that for low Ar pressures the Si layers are the predominant source of built-up stress. 
\end{abstract}
\maketitle

\newpage

Although adatom induced changes in surface facets of Si crystals has been much studied \cite{Robinson}, adatom induce changes in the surfaces of amorphous Si have not been reported, nor are we aware of such studies for the surface of other amorphous material. There are a wide variety of applications for amorphous thin films, and the atomic origins  of interface stress in thin films formed by sputter deposition bear on these applications \cite{Freund}. 

We report the observation of tension for the first  nanometer of WSi$_2$ deposited on amorphous Si, and these results are counter to the often invoked lock-down mechanism which results in compressive stress.  Sputtered multilayers are amorphous \cite{Moss},\cite{Laaziri} and  are ostensibly  free of crystalline relaxation mechanisms such as dislocation motion and crystalline anisotropies \cite{Mayr}. This makes them more ideal for the study of stress induced by local relaxation such as hybridization caused by adatoms.

As an example of a particular application, Si/WSi$_2$ multilayers consisting of many hundreds of  periods  have been used to make lenses for  nanofocusing of hard x-rays.\cite{KangPRL},\cite{KangAPL}.  Successful focusing lenses have been made in spite of the built-up stress , but the processing steps used  to slice out a  thin lens starting from a deposited  wafer can result in delamination  during mechanical polishing \cite{KangRSI}.  As a result of a grown-in stress,  multilayers  with an engineered reduction in accumulated  film  stress  will  likely  be  required  to  realize ready-to-use  Multilayer-Laue-Lens optics made  from 100 microns or more of  sputtered thickness. The present results indicate  a means to obviate some of the stress.

\begin{figure*}
\includegraphics{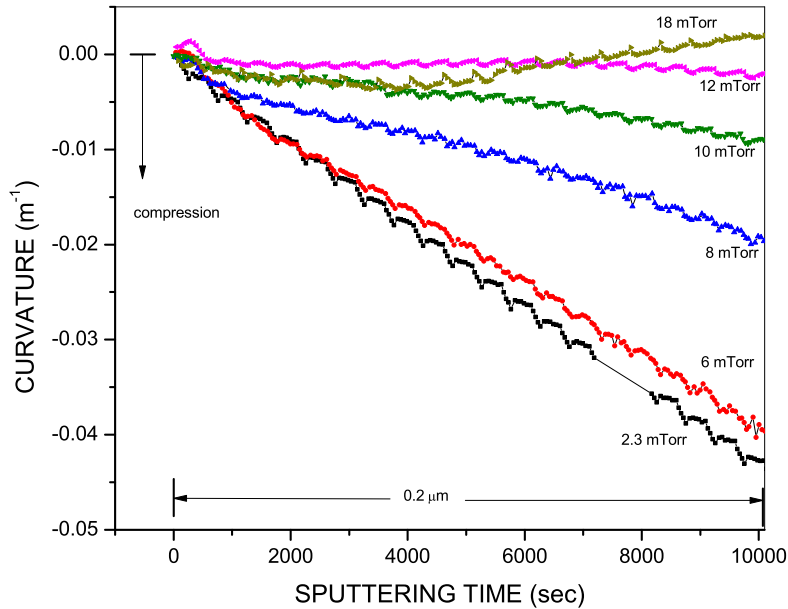}
\caption{\label{fig1} Curvature data for sputtering at  6  different Ar pressures for a total of 20 multilayer periods. Compressive changes are negative and tensile changes are positive. For the solid points, the most compression develops  for 2.3 mTorr, with progressively less compression for 6, 8, 10, and 12 mTorr. The data for 18 mTorr reveal a net tension after 20 periods.  (An unintended gap in the data  at 2.3 mTorr is also evident. Data in this range were not used in the analyses). (Color on-line only).}
\end{figure*}

We studied multilayers and made repeated curvature measurements for   twenty periods.  Measurements of the curvature obtained for 20 multilayer periods are shown in Fig.1.  

The thickness of both the Si and the WSi$_2$  layers were 5  nm, and 20 period multilayers were deposited onto Si(100) substrate wafers having a thickness of 262 $\mu$m. For sputter deposition the wafer was rotated to face either a WSi$_2$ or a Si target \cite{Conley}.  Ar plasma pressures of 2.3,  6,  8, 10, 12, and 18 mTorr were investigated.  After sputtering of 1  nm the wafer was rotated to face a laser based system for measurement of wafer curvature \cite{MOS}.  Stoney$^{\prime}$s equation in differential form  for the change in wafer curvature , $\delta(1/R)$, for a change in film thickness, $\delta(t_f)$ , is given by \cite{Stoney},
 
\begin{equation}
\delta(\frac{1}{R})=6\sigma(\frac{1-\nu}{E})\frac{\delta(t_f)}{t_s^2}
\label{eq:one}
\end{equation}

Here R is the radius of curvature of the wafer, $\sigma$ is the biaxial stress, $\nu$ is Poisson$^{\prime}$s ratio and E is the Young$^{\prime}$s modulus of the substrate, $t_s$ is the substrate thickness, and $t_f$ is the film thickness.  For  Si(100)  the ratio $(1-\nu)/E$ is conveniently isotropic in the plane of the wafer. Curvature measurements at azimuthal angles 90 degrees apart were found to give the same results which allowed us to rule out spurious effects due to mounting stresses or due to a flat on the otherwise round wafer.  Confirmation of the  multilayer period thickness was obtained with standard x-ray diffractometry.  

Intriguingly  and counter to the conventional expectation that a heavy  ion impinging on a lighter one should cause interface mixing, in previous synchrotron studies a smoothing effect was  instead found \cite{Wang} . The x-ray reflectivity measurements reported by Wang et al. \cite {Wang} were made in-situ and continuously as shutters to  sputtering guns were open and closed. The reflectivity was found to oscillate during growth, as expected from interference effects, but the reflectivity was constant when the shutters were closed. We infer that there is no change in either the roughness, density or  thickness of a grown layer  with a closed shutter, since changes in any of these would affect the reflected intensity. Consequently,  for the present experiments, possible   changes to the sputtered layers during the time that our samples were rotated to face to the MOS system are not considered significant. The results reported by Wang et al. \cite{Wang} are  that the  roughness builds up to be greater than or equal to 0.36 nm rms as each Si layer in a Si/WSi$_2$ multilayer  is grown,  but that each WSi$_2$ layer smoothens to a roughness  less than or equal to 0.27 nm rms . These results indicate that the surface morphology  of a sputtered Si layer is strongly affected by the deposition of WSi$_2$ ions. The present data add the information that this change in surface morphology results in a tension increment. 

Our data  reveal  a systematic change from a final compression at low Ar pressures to a final tension at higher pressures. This is a well known effect attributable to island coalescence and has been reported for a wide variety of systems (see Freund and Suresh, Fig. 1.39 \cite{Freund}).  In the present case  at Ar pressures above 6 mTorr, x-ray GISAX data has revealed  sputter deposition  by landing of coalesced WSi$_2$ particles, and this is consistent with an island coalescence model at higher Ar pressures \cite{Zhou}.

\begin{figure*} 
\includegraphics{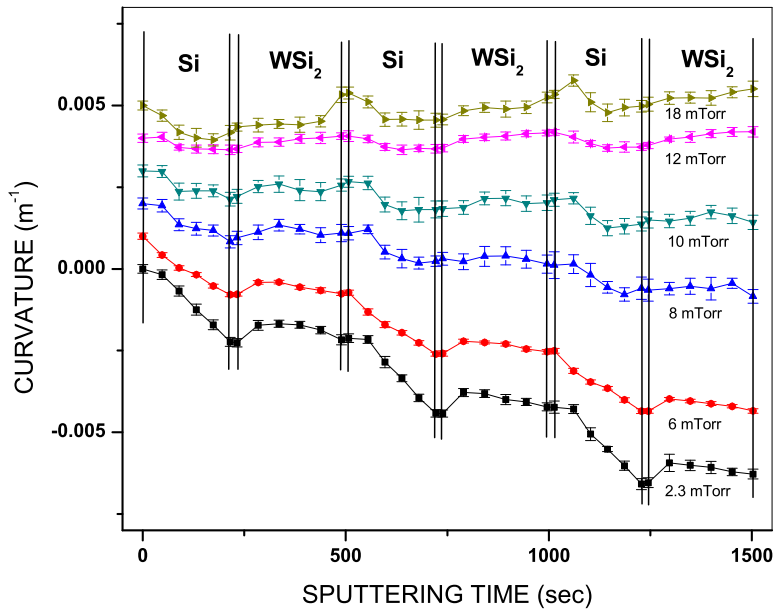}
\caption{\label{fig2} Curvature data as a function of growth time over three periods in the linear regime.   Positive changes correspond to tension and negative to compression. The thickness of the both  types of layers was 5 nm grown in 5 steps of 1 nm each. The growth rate of the WSi$_2$ was slightly less than for the Si with correspondingly slightly longer sputtering time  to achieve the same 5 nm thickness.  The data sets at each pressure have been arbitrarily offset for clarity.  Error bars represent ± one standard deviation for 18 curvature measurements made at each point.  (Color on-line only).}
\end{figure*}

\begin{figure*}
\includegraphics{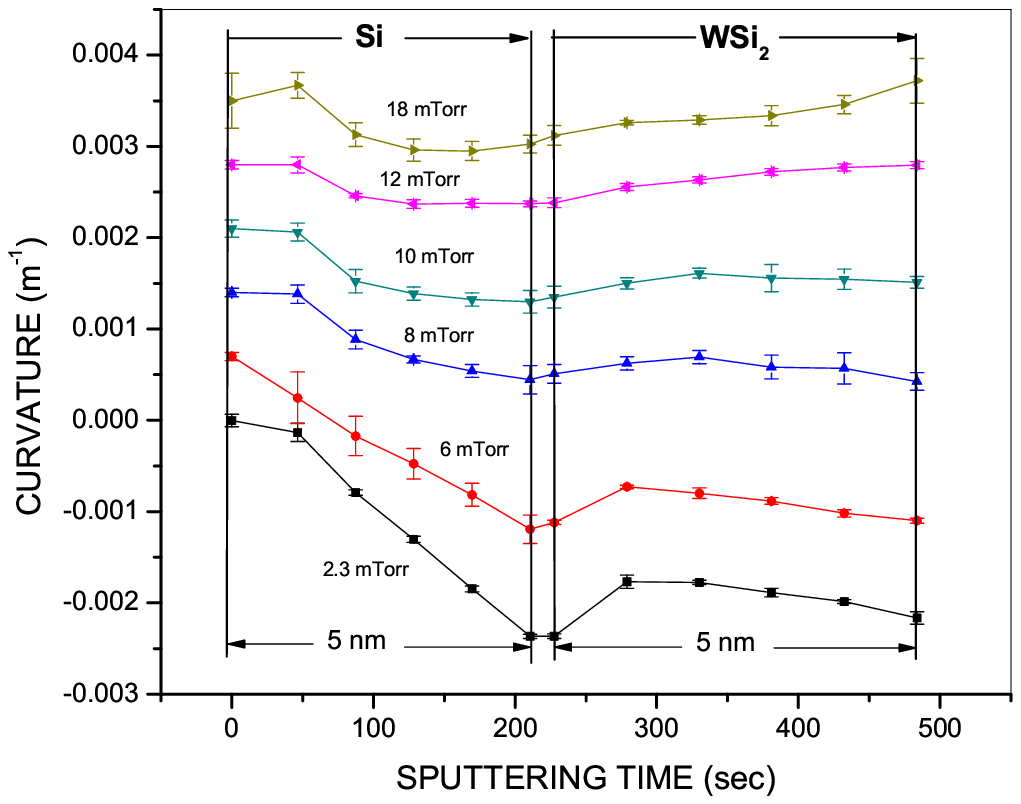}
\caption{\label{fig3} Curvature data that have been collapsed onto the time base for a single multilayer period and offset for clarity. Here the error bars correspond to plus and minus a standard deviation for  averaging  from 5 to 8 collapsed periods. (Color on-line only).} 
\end{figure*}

The data reveal an overall gross variation with time, and a variation about an  averaged curve that repeats over the multilayer period.  In each case there is a region of the gross variation which is linear as shown in Fig.2  for three periods. Although previous ex-situ studies in our laboratory hinted at this variation \cite{Liu},  we are not aware of other  reports of a  systematic curvature  variation over a multilayer period. A very pertinent aspect presently is that these data  permitted statistical averaging to reveal  weak stress effects arising from the deposition of only 1 nm of thickness. Since the data were very periodic, they could  be collapsed into a time base corresponding to a single multilayer period and averaged to yield the results shown in Fig.3.    
 
An  unexpected feature of these data is the observation of tension in the first nanometer  of WSi$_2$. As in the case of crystals, surface reconstructions on the surfaces of amorphous solids should be  driven by an inequality between surface stress, f, and excess surface free energy per unit area, $\gamma$ \cite{Mayr},\cite{Cammarata}. We conclude that $(f -  \gamma)$  is negative for  the  interface  created when  WSi$_2$  is sputter deposited onto sputtered  Si  leading  to a negative biaxial film stress, i.e., a tension, given by  $\sigma = (f - \gamma)/t_0$, where $t_0$ is the thickness of the surface layer. 

Meade and Vanderbilt  report strong tensile stresses for Si(111) surfaces covered by adatoms with  tensile stresses ranging from 1.18 eV  to 1.70 eV per  1 x 1  cell \cite{Meade}.  The effect of adatoms on a  Si(111) surface was calculated by them to make a  tensile stress contribution \cite{Vanderbilt}.   Our data permit a quantitative comparison to these results. For the observed  tension increment of 6 x 10$^{-4}$ m$^{-1}$ observed at 2.3 mTorr pressure after deposition of 1 nm of WSi$_2$, we can apply Eq. 1 to yield a value of  $\sigma$ = 1.1 x 10$^{10}$ dyne/cm$^2$  . Here we applied a value of   $(1- \nu)/E$ = 0.554 x 10$^{-12}$  cm$^2$/dyne , for the biaxial modulus of the Si(100) substrate. The product of this stress and the thickness increment of   1 nm leads to a value 0.36 eV  per  1 x 1 cell (surface area equal to 0.047 nm$^2$). This value is smaller than those calculated by Meade and Vanderbilt \cite{Meade}, and we propose that is due to  the  amorphous Si surface in our case. 
 
A   widely applicable model for the stress evolution of deposited layers is discussed by  Freund and Suresh \cite{Freund} , and involves  regimes of stress that change with growth thickness.  A compressive stage occurs at first, followed by a tension regime. For the first stage a "lock-down" mechanism has been invoked \cite{Freund}. The idea underlying this  mechanism  arises from the observation that the  bond length of a small particle locked to the substrate is smaller than the same material in bulk form. This brings about compression as the layer is filled in. 
 
Our  observations of tension in the first nanometer of  WSi$_2$  run counter to this model, and thus we are led to invoke the above model of   surface rearrangement for underlying Si layers.  We note that once the first nanometer is deposited, we do see a compressive regime as is commonly expected for a peening effect  for deposition of  films with higher atomic masses.  

As concerns a means to minimize the build up of accumulated stress, we propose that for  Si/WSi$_2$ multilayer applications that require a given  period, a reduced Si layer thickness should be effective.

In summary, an unexpected tension increment for the first nanometer of WSi$_2$ deposited onto Si in sputtered multilayers is reported. This tension increment is unusual in that a compression in the first stage of growth has been reported for a  variety of other systems. Because the  present data were obtained with multilayers, a powerful new statistical analyses is possible that provides high sensitivity to small changes in wafer curvature resulting from incremental stress. Unusual smoothing of a sputtered Si surface by sputter deposition of WSi$_2$ was previously reported \cite{Wang} indicative of Si surface rearrangement,  and such a model is invoked presently.  Published calculations of tension caused by adatoms for a Si(111) surface  can be compared quantitatively to the present case, and although similar in magnitude, our results reveal a smaller adatom induced tension for an amorphous Si surface.

We acknowledge the support for K. MacArthur provided by Prof. A. Genis at Northern Illinois University. We thank Prof. R. Headrick at the University of Vermont for comments. This work was supported by the U.S. Dept. of Energy, Office of Science, under Contract No. DE-AC-02-06CH11357.

$^{a}$ Present address: University of Tennessee, Dept. of Electrical Engineering and Computer Science, Knoxville, TN.

$^{b}$ Present address: National Synchrotron Light Source II, Brookhaven National Laboratory, Upton, NY.

$^{c}$ Corresponding author: macrander@aps.anl.gov

\end{document}